\documentclass[12pt]{article}
\usepackage{graphicx}

\newcommand{\bp}{\mbox{\boldmath $p$}}
\newcommand{\bq}{\mbox{\boldmath $q$}}

\def\Pom{{\bf I\!P}}

\def\lsim{\mathrel{\rlap{\lower4pt\hbox{\hskip1pt$\sim$}}
		\raise1pt\hbox{$<$}}}         
\def\gsim{\mathrel{\rlap{\lower4pt\hbox{\hskip1pt$\sim$}}
		\raise1pt\hbox{$>$}}}         
\def\Krakow{
Institute of Nuclear Physics, Polish Academy of Sciences, ul. Radzikowskiego 152, PL-31-342 Krak\'ow, Poland
}
\def\Rzeszow{Faculty of Mathematics and Natural Sciences, University of Rzesz\'ow, ul. Pigonia 1, PL-35-310 Rzesz\'ow, Poland}
\def\support{\footnote{Work supported by 
the Polish National Science Centre grant DEC-2014/15/B/ST2/02528.
}}
\def\antoni{\footnote{also at University of Rzesz\'ow}}

\def\Title#1{\begin{center} {\Large #1 } \end{center}}
\def\Author#1{\begin{center}{ \sc #1} \end{center}}
\def\Address#1{\begin{center}{ \it #1} \end{center}}

\newenvironment{Abstract}{\begin{quotation}  }{\end{quotation}}
\newenvironment{Presented}{\begin{quotation} \begin{center} 
             PRESENTED AT\end{center}\bigskip 
      \begin{center}\begin{large}}{\end{large}\end{center} \end{quotation}}
\def\Acknowledgements{\bigskip  \bigskip \begin{center} \begin{large}
             \bf ACKNOWLEDGEMENTS \end{large}\end{center}}

\def\beq{\begin{equation}}
\def\eeq#1{\label{#1}\end{equation}}
\def\eeqn{\end{equation}}

\def\beqa{\begin{eqnarray}}
\def\eeqa#1{\label{#1}\end{eqnarray}}
\def\eeqan{\end{eqnarray}}

\let\bar=\overbar

\def\L{{\cal L}}

\def\Dslash{\not{\hbox{\kern-4pt $D$}}}
\def\dslash{\not{\hbox{\kern-2pt $\del$}}}

\def\msb{{\bar{\ssstyle M \kern -1pt S}}}

\begin{document}
\begin{titlepage}

\vfill
\Title{
Photoproduction of $J/\psi$ with dissociation of protons
}
\vfill
\Author{Wolfgang Sch\"afer \support}
\Address{\Krakow}
\Author{Anna Cisek}
\Address{\Rzeszow}
\Author{Antoni Szczurek \antoni}
\Address{\Krakow}

\vfill
\begin{Abstract}
	We present the cross sections for both electromagnetic and diffractive dissociation of protons for semiexclusive production of 
	$J/\psi$ mesons in proton-proton collisions at the LHC.
	Differential distributions in missing mass ($M_X$), as well as distributions in rapidity and transverse momentum of the $J/\psi$ are 
	calculated for $\sqrt{s} = 7$ TeV and 13 TeV proton proton collisions.
	We compare the distributions for purely electromagnetic and purely diffractive proton excitations/dissociation. 
	We predict cross sections for electromagnetic and diffractive excitations of similar order of magnitude. 
\end{Abstract}
\vfill
\begin{Presented}
EDS Blois 2017, Prague, \\ Czech Republic, June 26-30, 2017
\end{Presented}
\vfill
\end{titlepage}
\def\thefootnote{\fnsymbol{footnote}}
\setcounter{footnote}{0}

\section{Introduction}

The exclusive vector meson $V = J/\psi,\psi',\Upsilon$ production in proton-proton collisions 
as measured by the LHCb collaboration \cite{Aaij:2013jxj} has recently 
attracted a lot of attention.

Indeed, it turns out, that at LHC-energies the exclusive production cross section
probes the $\gamma p \to Vp$ amplitude at higher energies than were previously
available, e.g. at HERA. 

\begin{figure}[htb]
	\centering
	\includegraphics[width=.2\textwidth]{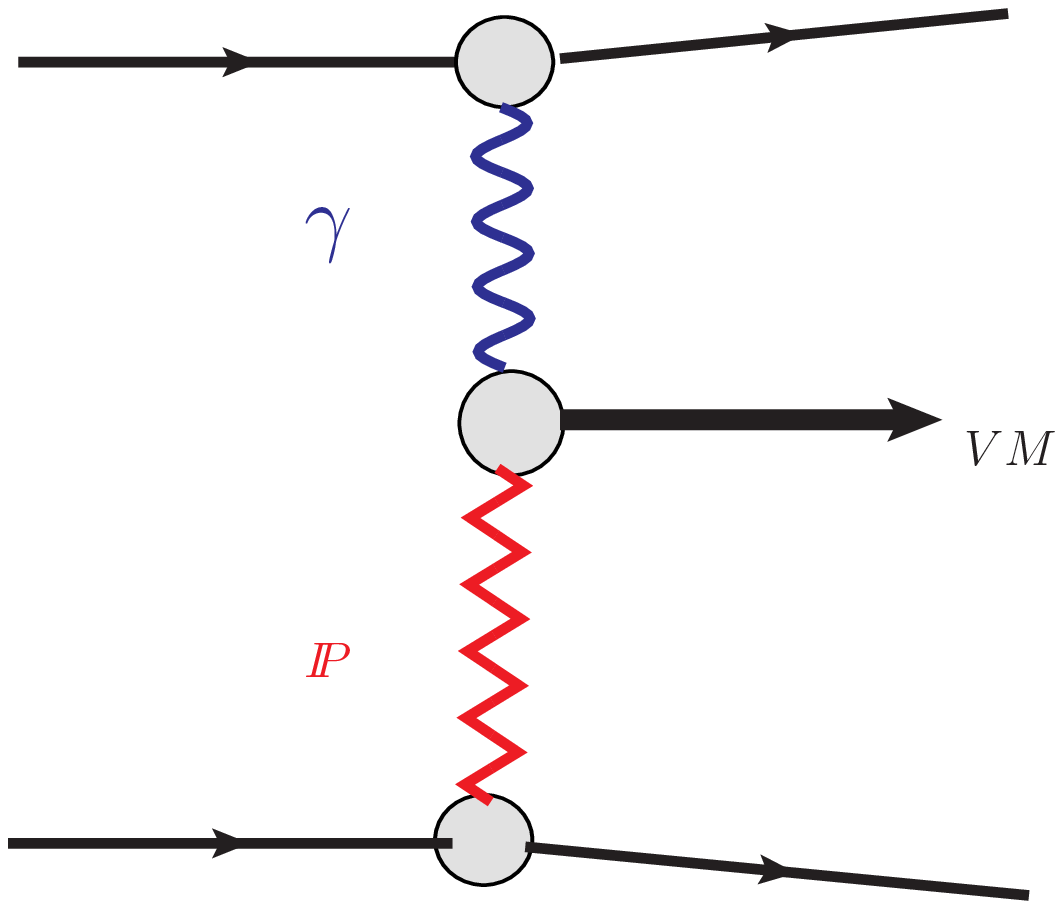}
	\includegraphics[width=.2\textwidth]{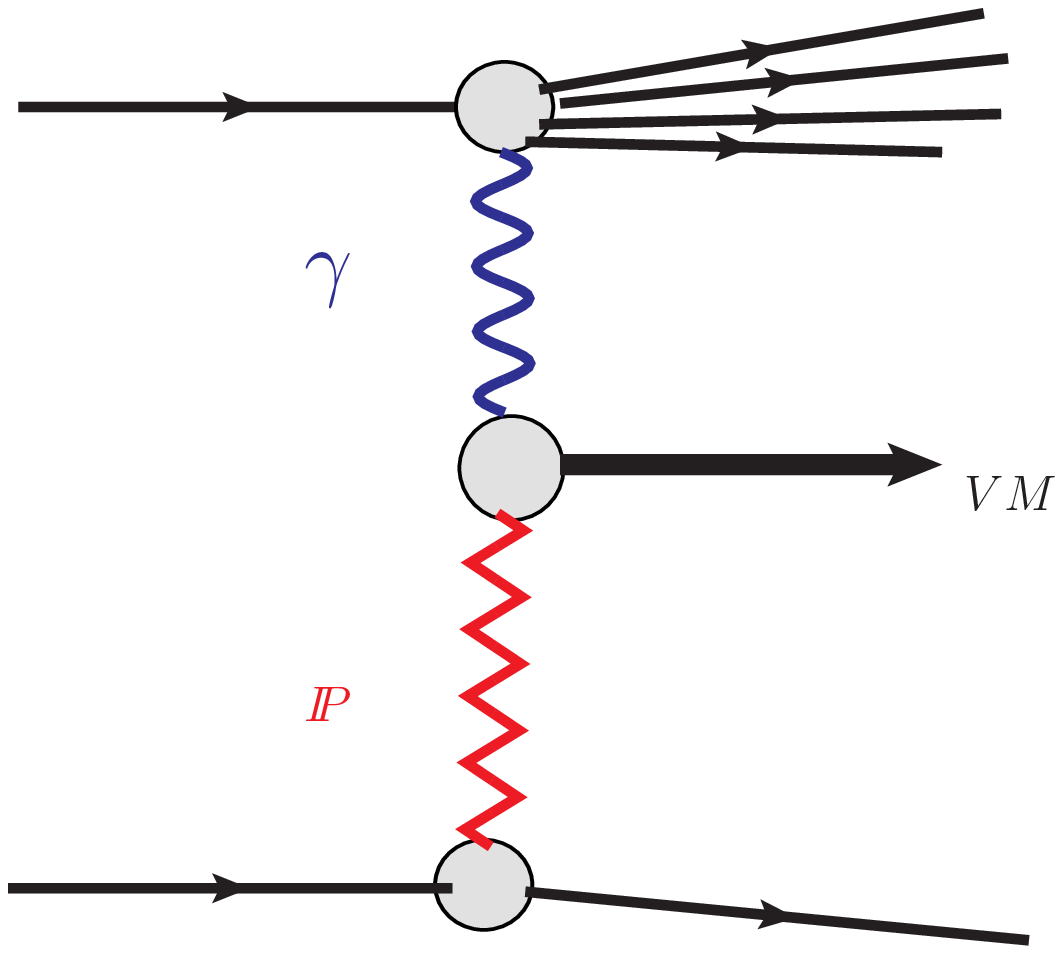}
	\includegraphics[width=.2\textwidth]{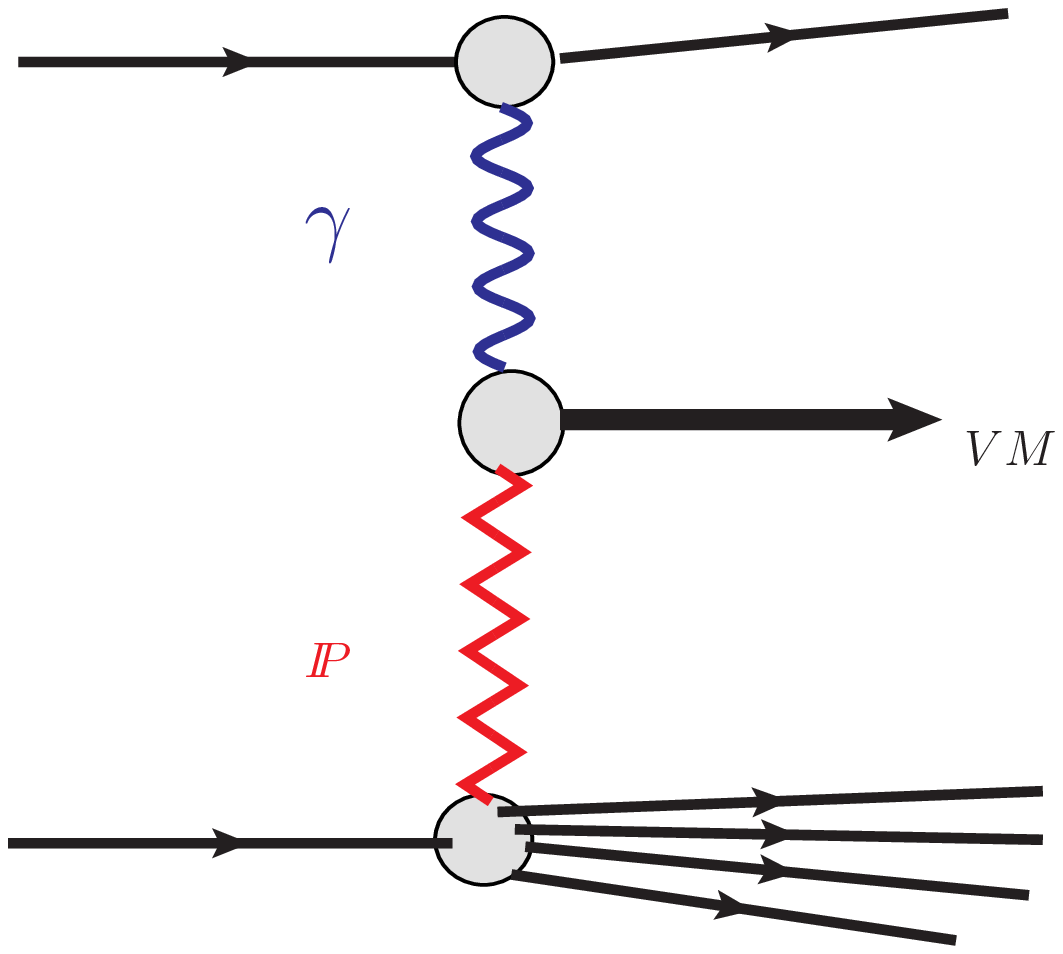}
	\caption{Born diagrams for three vector meson production mechanisms. Left panel: exclusive production through the $\gamma \Pom$-fusion,
		middle panel:  semiexclusive production with electromagnetic excitation of one of the protons, right panel: semiexclusive production 
		with diffractive dissociation of one of the incoming protons.}
	\label{fig_diag_1}
\end{figure}

For heavy vector mesons, the diffractive photoproduction amplitude in turn is a sensitive probe of the proton's
(unintegrated) gluon distribution (see e.g. the review \cite{Ivanov:2004ax} and references therein). 
A number of subtleties have to be taken into account, even in the case of fully exclusive production,
such as the interference of contributions where different protons emitted the photon and absorptive
corrections, all treated in \cite{Schafer:2007mm}.   

Here we are concerned with a different complication: due to the fact that up to now there are no measurements
with tagged protons, there is always a contribution with dissociation, which needs to be modelled.
As shown in fig.\ref{fig_diag_1}, the dissociation can be induced by electromagnetic as well as strong
interactions (diffractive dissociation). More details can be found in \cite{Cisek:2016kvr}.

\section{Diffractive photoproduction with electromagnetic dissociation}

The electromagnetic dissociation processes have a nice property, that they can be calculated in an 
entirely data-driven way \cite{daSilveira:2014jla,Luszczak:2015aoa}. 
In fact the $\gamma^* p \to X$ vertices can after summing over states $X$ be
related to the nucleon deep-inelastic structure function. 
The cross section for semiexclusive vector meson production then is given as 
\begin{eqnarray}
{d \sigma (pp \to X V p; s) \over dy d^2\bp} = 
\int {d^2\bq \over \pi \bq^2} {\cal{F}}^{(\mathrm{inel})}_{\gamma/p}(z_+,\bq^2) 
{1\over \pi} {d \sigma^{\gamma^* p \to Vp} \over dt}(z_+s,t = -(\bq - \bp)^2) +( z_+ \leftrightarrow z_-) .
\end{eqnarray}
Photons carry a longitudinal momentum fraction $z_\pm = e^{\pm y} \sqrt{\bp^2 + m_V^2}/\sqrt{s}$
and transverse momentum $\bq$. The effective photon flux in dissociative events is
\begin{eqnarray}
{\cal{F}}^{(\mathrm{inel})}_{\gamma/p}(z,\bq^2) = {\alpha_{\mathrm{em}} \over \pi} (1 - z) \int^\infty_{M^2_{\mathrm{thr}}} 
{dM_X^2 F_2(x_{\rm Bj},Q^2)  \over M_X^2 + Q^2 - m_p^2}  \Big[ {\bq^2 \over \bq^2 + z (M_X^2 - m_p^2) + z^2 m_p^2} \Big]^2 
\, ,
\end{eqnarray}
with
\begin{eqnarray}
Q^2 =  {1 \over 1 - z} \Big[ \bq^2 + z (M_X^2 -m_p^2)  + z^2 m_p^2 \Big], x_{\rm Bj} = {Q^2 \over Q^2 + M_X^2 - m_p^2} . 
\end{eqnarray}
In practical calculations, fits of $F_2$ from refs.\cite{Fiore:2002re,ALLM} have been useful.

\section{Diffractive photoproduction with strong dissociation}

For the description of diffractive dissociation, unfortunately we do not have much data to guide 
our calculations. We take two types of mechanisms into account. 
At low transverse momentum transfers, the excitation of resonances dominates. 
Due to the vacuum quantum numbers of the Pomeron these resonances
must be isospin $1/2$. Their spin and parity can however differ from the nucleon quantum numbers.
A model of \cite{JKLMO2011} includes resonances on the nucleon trajectory, $N^*(1680)$, $J^P={5 \over 2}^+$,
$N^*(2220)$, $J^P={9 \over 2}^+$, and $N^*(2700)$, $J^P={13 \over 2}^+$. 
The lowest-lying positive parity excitation, the Roper resonance $N^*(1440)$ is also included but does not play
a very important role. It should be pointed out, that negative parity resonances 
like $N^*(1520)$, $J^P = {3 \over 2}^-$ can also be diffractively excited, but are up to now not included 
in the model.

Large mass continuum dissociation at large $p_T$ is treated in a similar fashion as incoherent diffraction on a nucleus: 
we simply assume that the diffractive production takes place on a constituent of the target (a quark or gluon parton) 
and sum incoherently over partons, at a hard scale which corresponds to $p_T^2 + M_{J/\psi}^2$. 
See \cite{Cisek:2016kvr} for more information.


\section{Results and summary}

\begin{figure}[htb]
	\centering
	\includegraphics[width=0.4\textwidth]{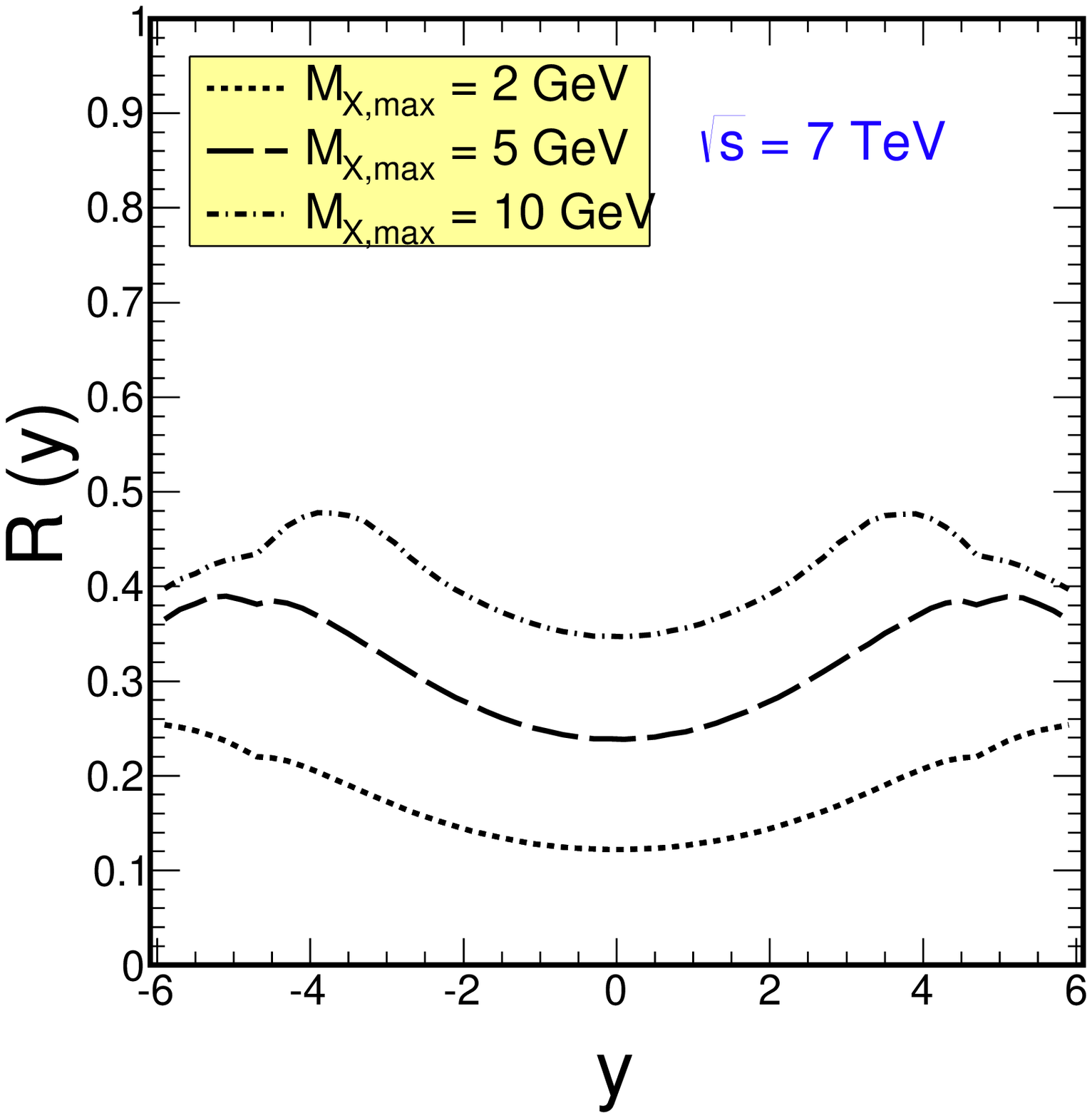}
	\includegraphics[width=0.4\textwidth]{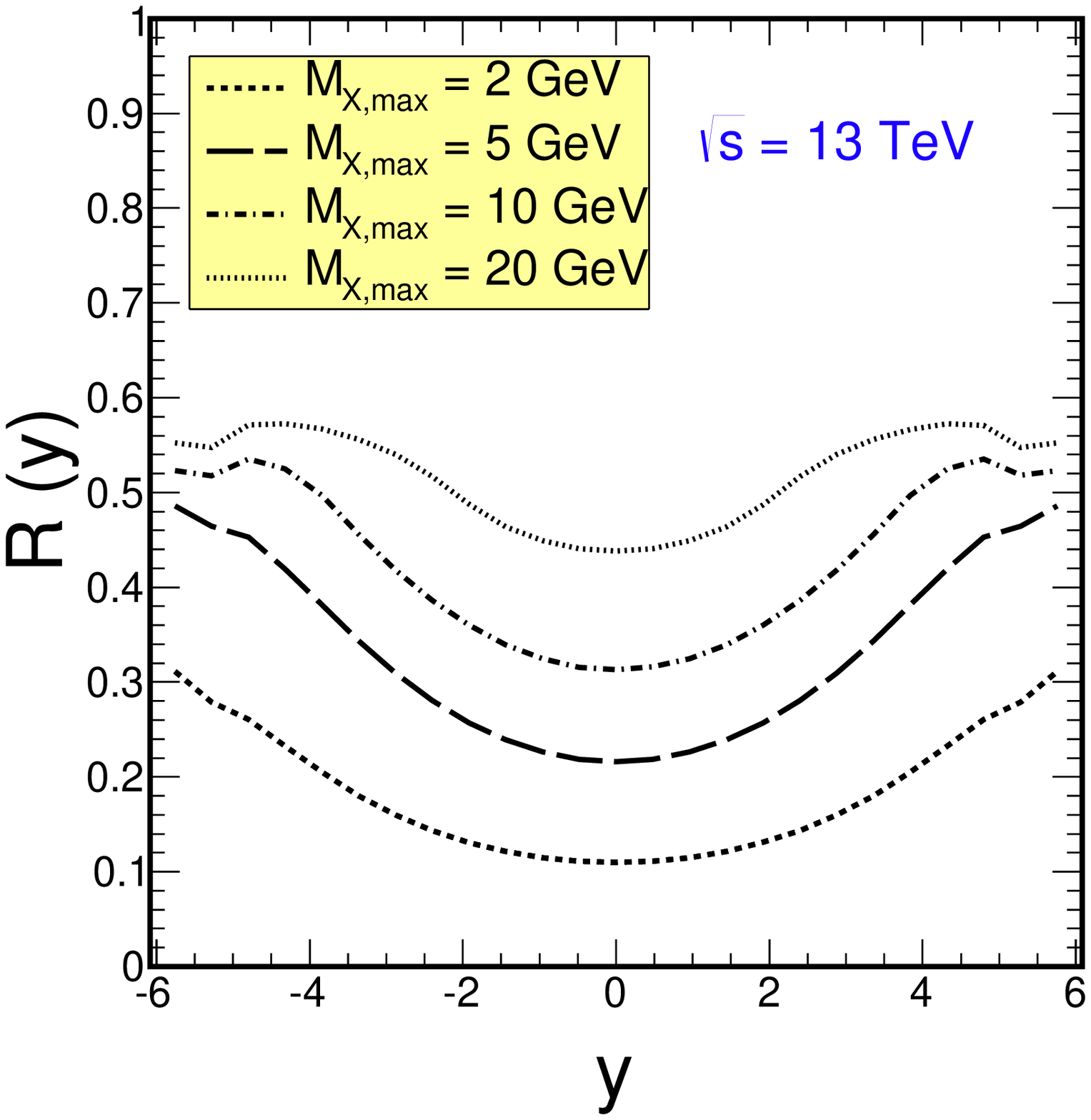}
	\caption{The ratio $R(y)$ of dissociative to exclusive production of $J/\psi$ as a 
	function of rapidity of  the meson.}
	\label{fig_dsig_dy}
\end{figure}

Let us present some numerical results \cite{Cisek:2016kvr} obtained from the approach described above.
In fig.\ref{fig_dsig_dy} we show the ratio of dissociative over exclusive $J/\psi$ 
production 
\begin{eqnarray}
R(y) &=& \frac{d\sigma_{p p \to p J/\psi X}(M_X < M_{X,\rm{max}})/dy}
{d\sigma_{p p \to p J/\psi p}/dy} \; . 
\nonumber
\end{eqnarray}
Here the dissociative cross section has been integrated up to $M_{X,\rm{max}}$, and results for 
several values of the upper limit are shown.  We also show two different $pp$ cms-energies.

\begin{figure}[htb]
	\centering
	\includegraphics[width=.3\textwidth]{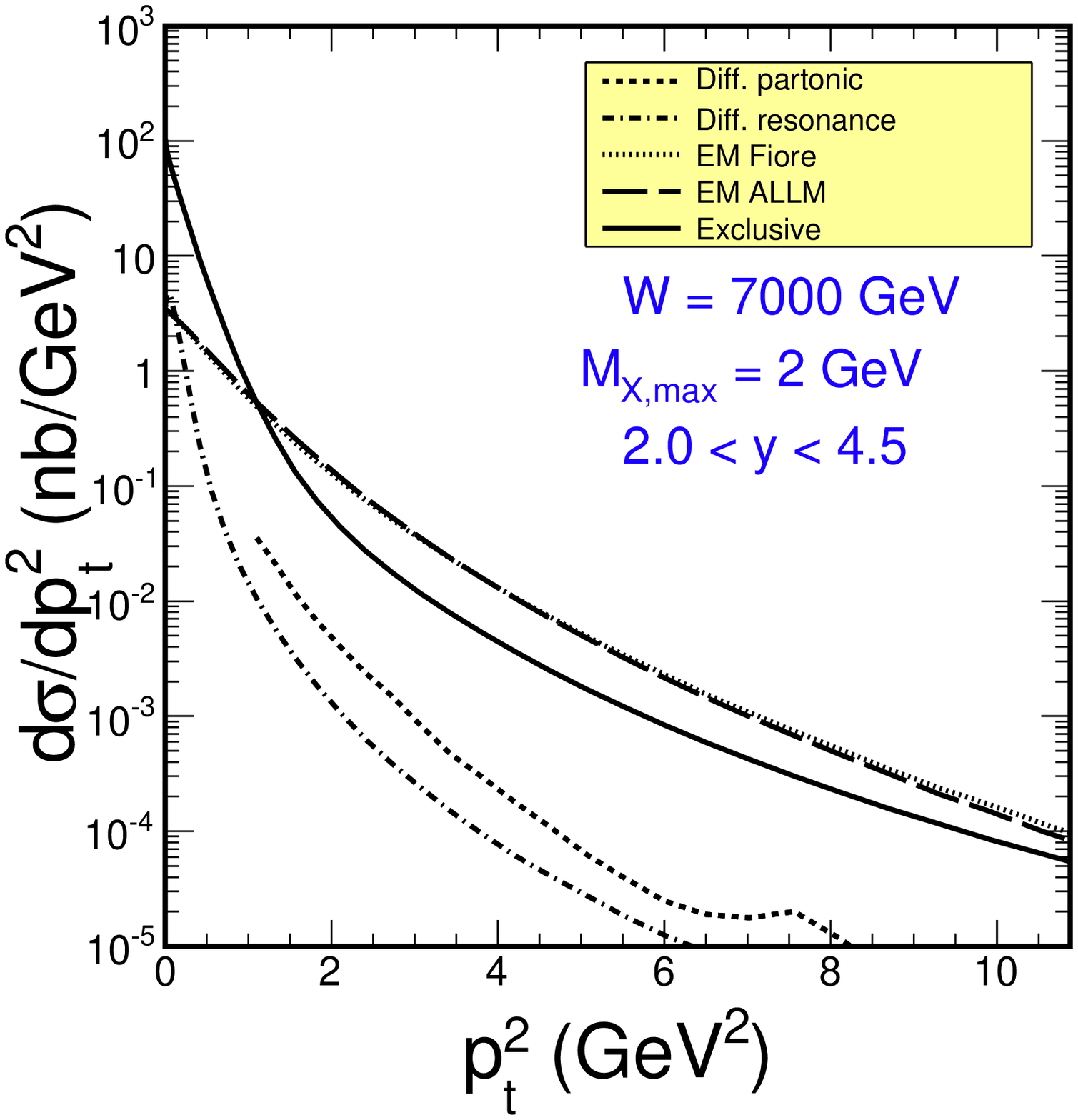}
	\includegraphics[width=.3\textwidth]{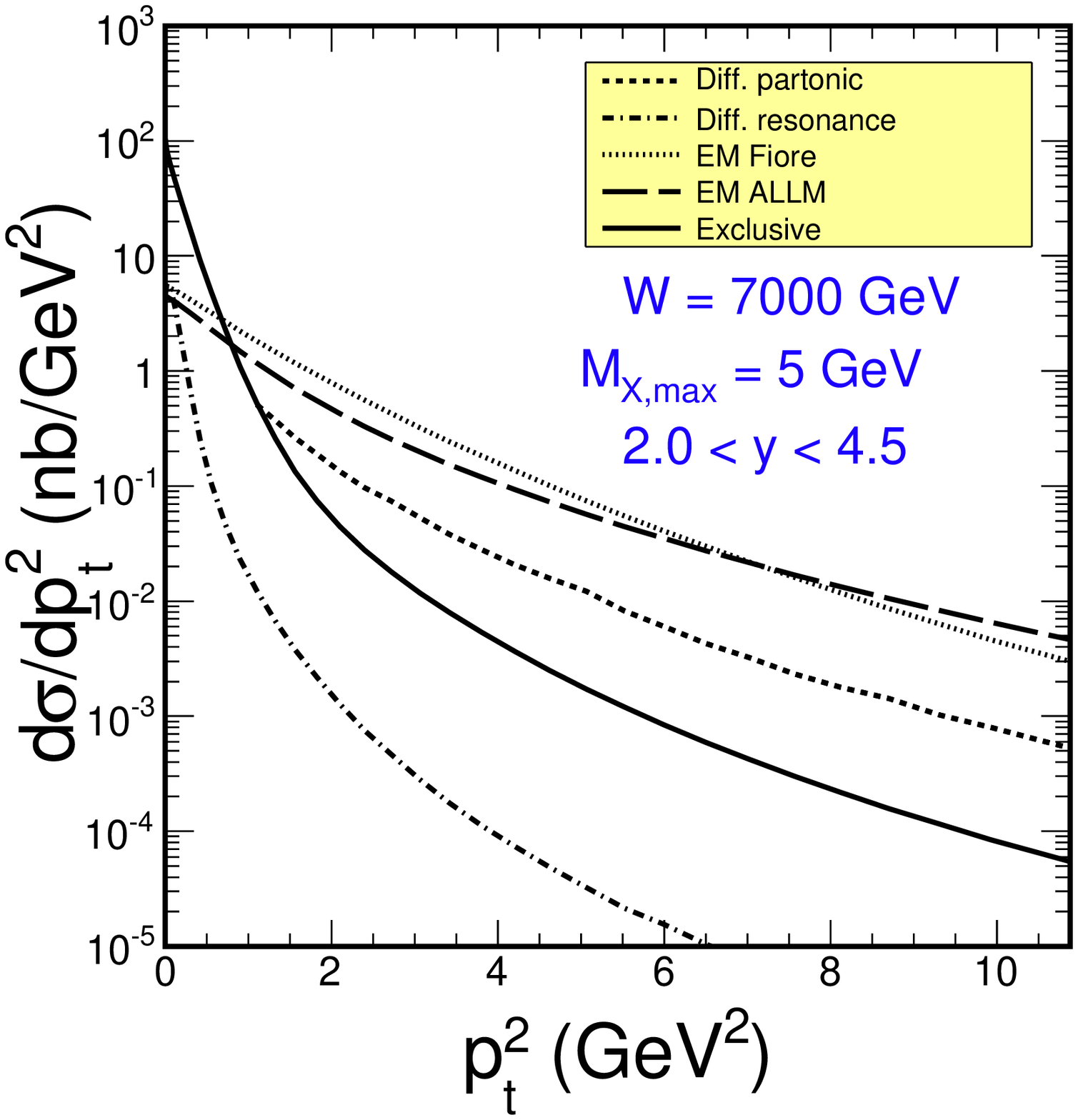}
	\includegraphics[width=.3\textwidth]{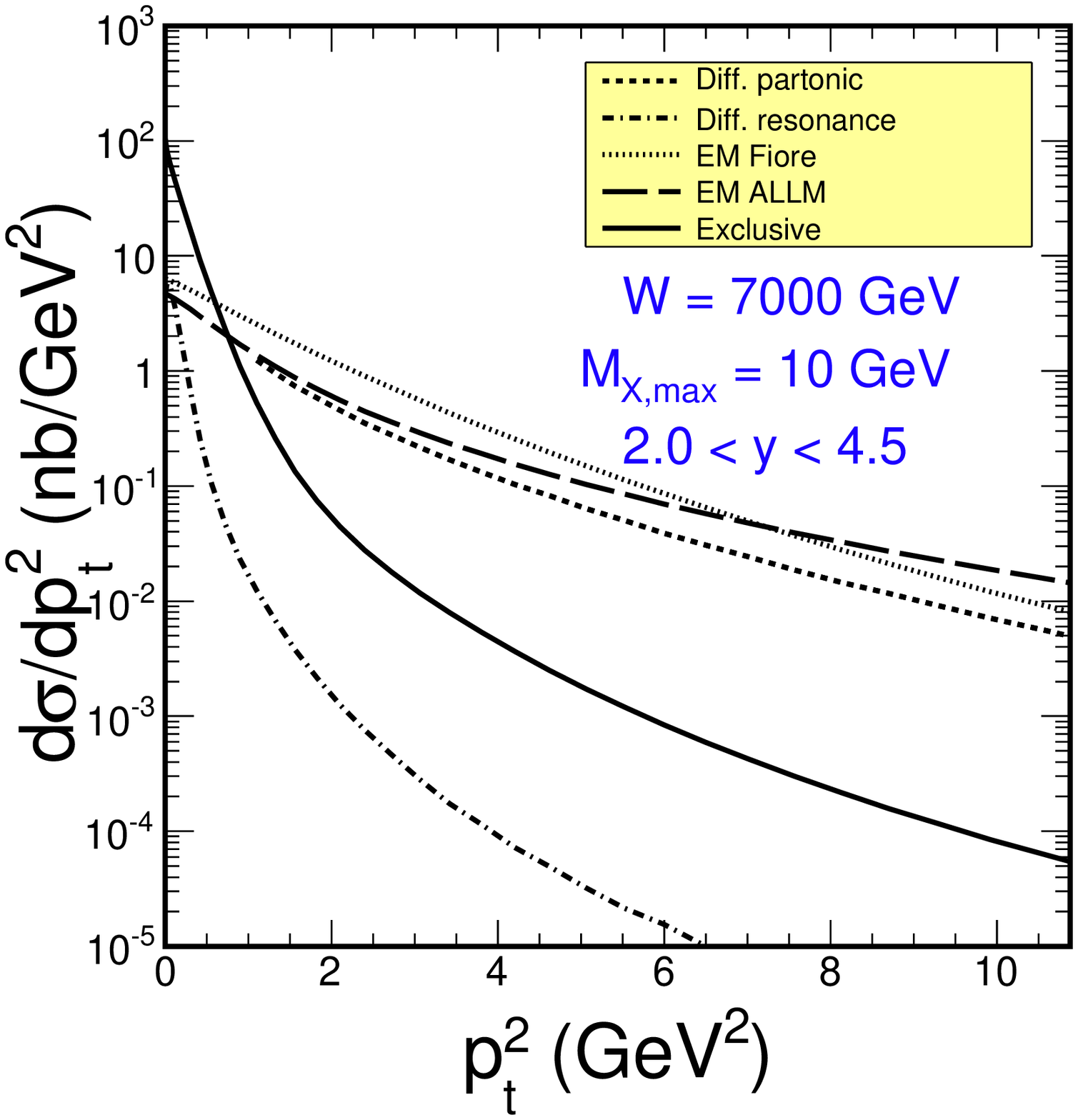}
	\caption{The differential $p_T^2$-distribution of $J/\psi$-mesons for different production mechanisms.
		The cuts in rapidity correspond to the range of the LHCb experiment. From left to right, the upper limit
	on the $M_X$ integration is changed.}
	\label{fig_dsig_dpt2}
\end{figure}

In fig.\ref{fig_dsig_dpt2} we show the distribution of the $J/\psi$ in $p_T^2$ for different production mechanisms.
We imposed cuts in rapidity that correspond to the ones of the LHCb experiment.
Exclusive production shows the typical coherent peak, while inelastic processes give rise to a much smaller slope
of the $p_T^2$ spectrum at larger $p_T^2$. We also observe, that electromagnetic dissociation is as important, or,
depending on the cuts on $M_X$, even more important than diffractive dissociation.

Finally, it should be mentioned that a measurement of the $J/\psi$ photoproduction at large $p_T$, but at
the same time with large rapidity gaps between the vector meson and other particles is interesting in
its own right. Similarly to the gap-jet-gap cross section it can contain useful information 
on the perturbative QCD Pomeron.

\Acknowledgements 

The work presented here was partially supported by the 
Polish National Science Centre grant DEC-2014/15/B/ST2/02528 and 
by the Center for Innovation and Transfer of Natural Sciences and 
Engineering Knowledge in Rzesz\'ow.

\end{document}